\documentstyle[psfig]{kapproc}
\kluwerbib
\startauthorindex
\begin{document}
\articletitle{Solar-Type Post-T Tauri Stars in the Nearest OB Subgroups}
\articlesubtitle{}
\author{Eric E. Mamajek}
\affil{Steward Observatory, The University of Arizona,\\
933 N. Cherry Ave., Tucson AZ 85721}
\email{eem@as.arizona.edu}
\chaptitlerunninghead{Sco-Cen Post-T Tauri Stars}
\anxx{Mamajek}
\begin{abstract}
I discuss results from the recent spectroscopic survey 
for solar-type pre-MS stars in the Lower Centaurus-Crux (LCC) 
and Upper Centaurus-Lupus (UCL) OB subgroups by 
Mamajek, Meyer, \& Liebert (2002, AJ, 124, 1670). LCC and UCL
are subgroups of the Sco-Cen OB association, and the
two nearest OB subgroups to the Sun. In the entire survey of
110 pre-main sequence stars, there exists only one 
Classical T Tauri star (PDS 66), implying that only $\sim$1\% of $\sim$1 
M$_{\odot}$ stars are still accreting 
at age 13$\pm$7 (1$\sigma$) Myr. Accounting for observational errors, 
the HRD placement of the pre-MS stars is consistent with the bulk of 
star-formation taking place within 5-10 Myr. 
In this contribution, I estimate conservative 
upper limits to the intrinsic velocity 
dispersions of the post-T Tauri
stars in the LCC and UCL subgroups ($<$1.6 km\,s$^{-1}$ and
$<$2.2 km\,s$^{-1}$, respectively; 95\% CL) 
using Monte Carlo 
simulations of Tycho-2 proper motions for candidate subgroup members.
I also demonstrate that a new OB subgroup recently proposed to exist 
in Chamaeleon probably does not.
\end{abstract}

\section*{Introduction}
In understanding the development of our own solar system and the 
formation of stars and planets in general, we would like to know: 
How long does star-formation persist in molecular clouds?
How long do stars accrete from circumstellar disks? 
What controls the rotational evolution of pre-main sequence stars? 
What are the characteristics and frequency of dusty debris disks 
around solar-type stars? 
General questions regarding star and planet formation can be 
addressed by identifying and investigating large samples of 
pre-main sequence stars in the nearest OB associations.
Understanding the evolution of low-mass stars
intermediate in age between embedded T Tauri stars (TTSs) and zero-age 
main sequence (ZAMS) stars has been historically impeded by the lack of 
appropriate stellar
samples. T Tauri stars ($\leq$few Myr-old) are found in great numbers in and 
near molecular clouds, while well-characterized ZAMS stars ($\sim$30-100 
Myr) are abundant in nearby open clusters. Finding the elusive, 
intermediate-age, pre-MS ``post-T Tauri'' stars has recently become 
possible due to the availability of the {\it ROSAT} All-Sky Survey 
and {\it Hipparcos}/Tycho databases (Jensen 2001).\\

In this contribution, I summarize the findings of a recent 
spectroscopic survey of post-T Tauri stars in the 
Lower Centaurus-Crux (LCC) and Upper Centaurus-Lupus 
(UCL) subgroups (Mamajek, Meyer, \& Liebert, 2002; hereafter MML02).
These are the two oldest subgroups of the Sco-Cen OB association,
and the two nearest OB subgroups to the Sun. 
I also calculate an upper limit to the 
velocity dispersion of low-mass members of LCC and UCL, and 
critically examine the evidence for a new OB subgroup in 
Chamaeleon.

\section{Post-T Tauri Stars in LCC and UCL}

MML02 conducted a spectroscopic survey of a proper motion-
and X-ray-selected sample of stars in the LCC and UCL regions of the 
Sco-Cen OB association. For the survey, MML02 used the Astrographic 
Catalog-Tycho (ACT) and Tycho Reference Catalog (TRC) astrometric catalogs and
the {\it ROSAT} All-Sky Survey Bright Source Catalog (RASS-BSC) of X-ray
sources. The proper motion candidates
were selected by Hoogerwerf (2000) and MML02 as being stars whose both
ACT and TRC proper motions were consistent with LCC or UCL membership, and
whose B-V colors and V magnitudes were between 3 mag above and 1 mag below
the Schmidt-Kaler (1982) ZAMS at the mean distance for each OB subgroup. To 
narrow our selection of young stellar candidates, we observed only those
stars that had X-ray sources within 40'' radius in the RASS-BSC.
We added to our target list the G and K-type Hipparcos stars selected by
de Zeeuw et al (1999) as probable LCC and UCL members. 
Blue and red medium resolution spectra were taken with
the Dual-Beam Spectrograph on the Siding Springs 2.3-m telescope in 
April 2000. From the proper motion- and X-ray-selected sample, we identified the 
pre-MS stars 
spectroscopically through the following criteria: late spectral types (FGK),
Li-rich (Li I $\lambda$6707 line), subgiant surface gravities (using a
band-ratio measurement of the Sr II $\lambda$4077 and Fe I $\lambda$4071 
absorption lines), and HRD positions above the main sequence. The success rate
for detecting Li-rich subgiants (i.e. probable pre-MS stars) among
the proper motion- and X-ray selected sample was 93\%, compared
to 73\% for the de Zeeuw et al. (1999) kinematic sample. 
Only one star in the sample (MML 34 = PDS 66) had strong H$\alpha$ emission
and a statistically-significant K-band excess consistent with being
a Classical T Tauri star. The MML02 survey demonstrated the following: 

\begin{itemize}

\item The mean pre-MS isochronal ages of LCC and UCL are nearly identical, 
and agree well with new estimates of the turn-off ages ($\sim$15-23 Myr, 
depending on choice of evolutionary tracks).

\item Only $\sim$1\% of solar-type stars in our sample are Classical
T Tauri stars. The mean age for the sample, biased toward younger ages 
for the lower mass stars, is 13 Myr using the D'Antona \& Mazzitelli (1997) 
tracks. The incidence of accretion disks is consistent with the
idea that accretion terminates in solar-type stars 
within a $\sim$10 Myr timescale.

\item The band ratio Sr II $\lambda$4077/Fe I $\lambda$4071 is very
useful for segregating Li-rich stars into pre-MS and ZAMS stars. This
band-ratio defines clear loci for dwarfs and subgiants among
spectral standards.

\item 95\% of the low-mass star-formation in each OB subgroup must have
taken place within a 8-12 Myr span. 
 
\end{itemize}

\section{The Velocity Dispersion of the LCC and UCL Post-T Tauri Stars}

With a new, high-quality astrometric catalog now available (Tycho-2), 
one can address the 
question: is the internal velocity dispersion of the post-T Tauri 
members the same as that for the high mass members ($<$1-1.5\,km\,s$^{-1}$; 
de Bruijne 1999a)? Is it measurable with existing data? Our group has taken 
echelle spectra of all of the
MML02 pre-MS stars, with one goal being to measure the velocity dispersion
and possible expansion of the OB subgroups. One can, however, calculate an
upper limit to the velocity dispersion with the Tycho-2 
astrometry alone.\\

Pre-main sequence members of the OB subgroups can be efficiently
selected by their strong X-ray emission and convergent proper motions
(MML02). To identify low-mass member candidates,
I construct a cross-referenced catalog of all Tycho-2 stars
with {\it ROSAT} 
All-sky Survey BSC and FSC X-ray sources within 40'' radius (hereafter
RASS-TYC2), and analyze the
distribution of their proper motions. 
Within the LCC and UCL regions (defined by de Zeeuw et al. 1999), I find 
271 RASS-TYC2 stars with (B-V)$_{J}$ $>$ 0.60 (G-type or later) in LCC and 
328 in UCL. For simplicity, I do not apply a magnitude restriction other than 
the Tycho-2 magnitude limit -- the vast majority are consistent with being 
pre-MS or ZAMS at $d$ = 100-200 pc.\\ 

To search for subgroup members, I plot the proper motions for the RASS-TYC2 
stars in ($\mu_{\upsilon}$, $\mu_{\tau}$) space instead of 
($\mu_{\alpha}$, $\mu_{\delta}$). The proper motion components 
represent the motion toward the subgroup convergent 
point ($\mu_{\upsilon}$) and perpendicular to the great circle between the star 
and the convergent point ($\mu_{\tau}$) (Smart 1968). 
The expectation value of $\mu_{\tau}$ 
for an ensemble of bona fide cluster members is zero, 
and $\mu_{\upsilon}$ scales with distance
and angular separation from the convergent point.  
I adopt the space motion for the LCC and UCL subgroups from de Bruijne (1999a) 
(the $g_{lim}$ = 9 solutions in their Table 5, where the space motion vector
can be converted to a convergent point solution using eqn. 10 of
de Bruijne (1999b)).
Low-mass members of the OB subgroups stand out clearly in ($\mu_{\upsilon}$, 
$\mu_{\tau}$) space (Fig. 1). The mean subgroup proper motion values 
(de Bruijne 1999a; Table 4) are shown as dashed vertical lines, where $\bar{\mu}$ for 
members is approximately equal 
to $\bar{\mu}_{\upsilon}$. I define boxes around the loci in Fig. 1
to select probable association members for statistical study
(70 stars in LCC, 105 in UCL).\\

\begin{figure}[ht]
\sidebyside
{\centerline{\psfig{file=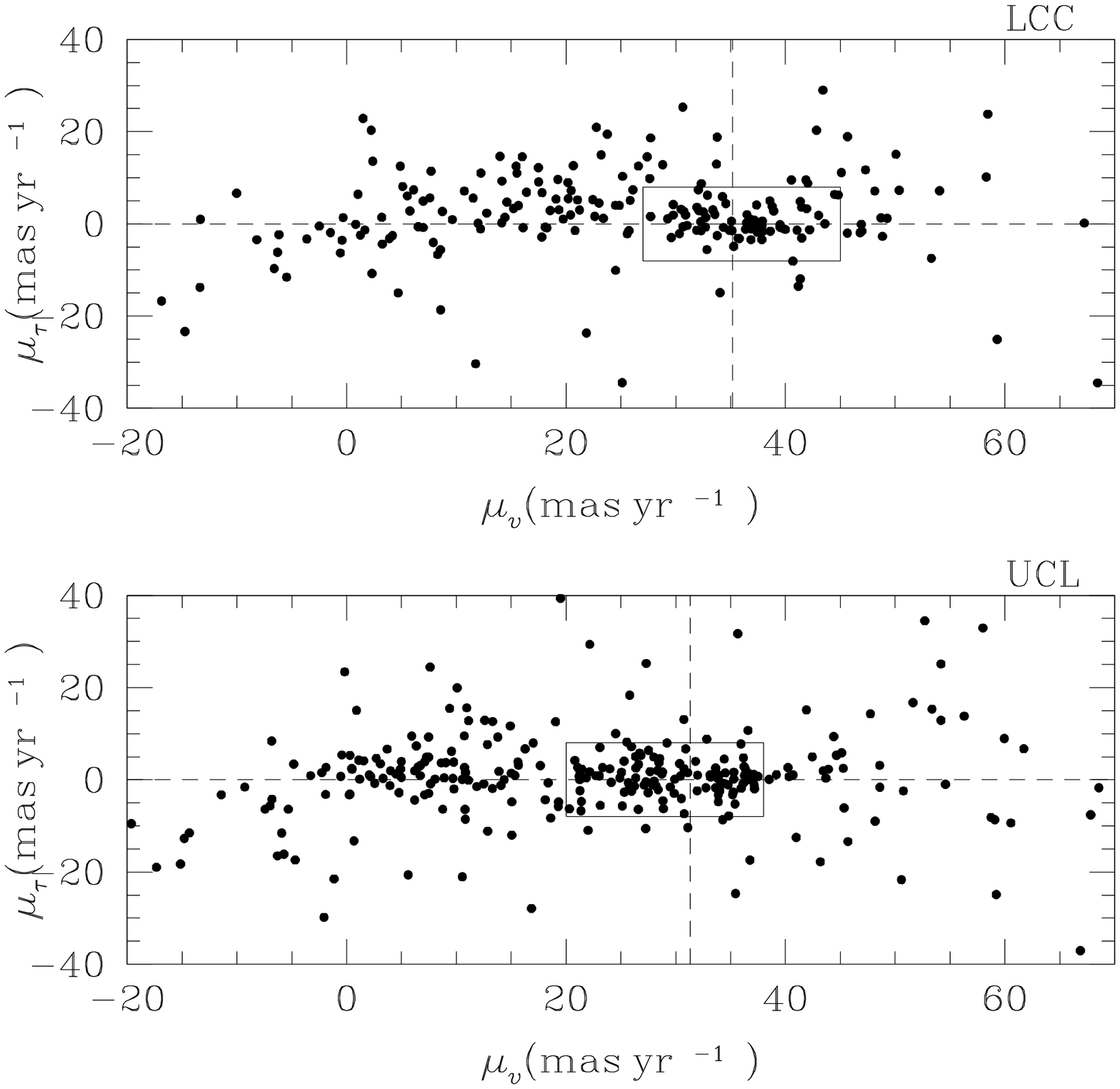,height=2.3in}}
\caption{Proper motions of RASS-TYC2 stars with (B-V)$_{J}$ $>$ 0.60, 
lying within the sky regions of the LCC and UCL OB subgroups defined 
by de Zeeuw et al. (1999). The mean $\bar{\mu}_{\upsilon}$ values
for the OB subgroups are shown by vertical dashed line (from
de Bruijne 2000). Association members should have a mean value of 
$\bar{\mu}_{\tau}$ = 0, with small scatter $\sigma$($\mu_{\tau}$).}}
{\centerline{\psfig{file=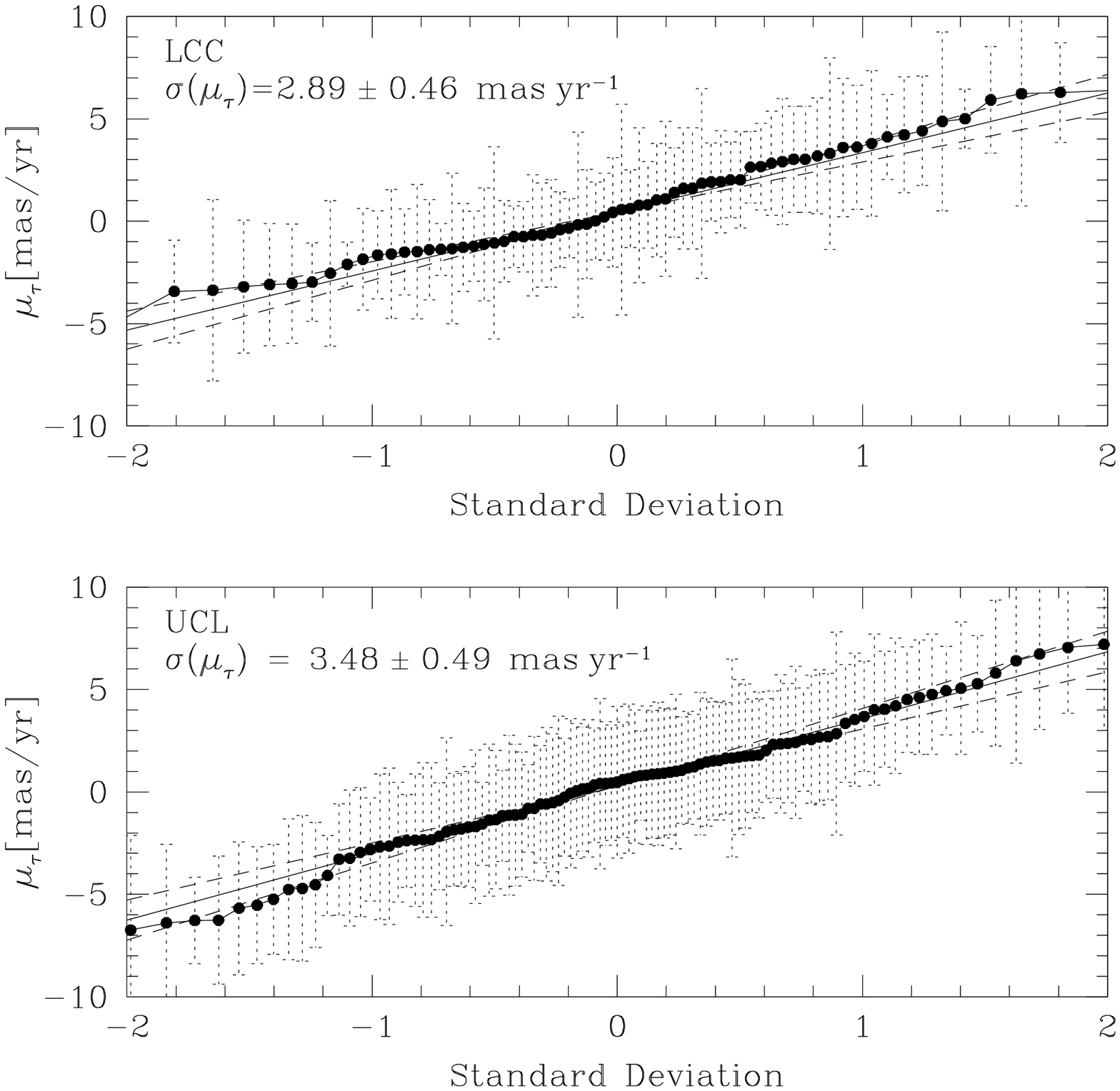,height=2.3in}}
\caption{Probability plots for $\mu_{\tau}$
for stars in the boxes defined in Fig. 1. A Gaussian
profile will produce a straight line, where the slope gives
the standard deviation. 
The observed distributions are consistent with
the Tycho-2 proper motion errors combined with intrinsic velocity dispersions
of 0.6$^{+0.5}_{-0.6}$ km\,s$^{-1}$ (LCC) and 
1.0$^{+0.6}_{-1.0}$ km\,s$^{-1}$ (UCL).}}
\end{figure}
\noindent

In order to estimate the observed dispersion in $\mu_{\tau}$ in a 
way that is insensitive to the boundaries of the subjectively
drawn selection box (Fig. 1), and the presense of outliers, 
I use probability plots (Fig. 2). I follow the analysis method
outlined in \S3 of Lutz \& Upgren (1980).
In a probability (or ``probit'') plot, 
the abscissa is the expected deviation from the
mean predicted for the $i$th sorted data point in units 
of the standard deviation,
and the ordinate is the data value in question ($\mu_{\tau}$).
The slope of the probability plot distribution yields 
the standard deviation, and the $y$-intercept is the median. 
To fit a line to the probability plots in Fig. 2, I use the
Numerical Recipes least-squares routine {\it fit}, and trim
10\% from both sides of the distribution to mitigate against the
effects of outliers. 
The probability plots yield observed standard deviations of 
$\sigma(\mu_\tau)$ = 2.9\,$\pm$\,0.5 mas\,yr$^{-1}$ (LCC) and 
3.5\,$\pm$\,0.5 mas\,yr$^{-1}$ (UCL). 
The mean values of $\mu_{\tau}$ are close to zero (0.7\,$\pm$\,0.3 
mas\,yr$^{-1}$ for LCC, 0.3\,$\pm$\,0.3 mas\,yr$^{-1}$ for UCL), 
consistent with the expectation that most of the RASS-TYC2 stars
in the boxes in Fig. 1 are subgroup members.\\

An upper limit to the velocity dispersions of the low-mass
subgroup memberships can be estimated as follows. The observational errors  
in $\mu_{\tau}$ range widely from 1-7 mas\,yr$^{-1}$ 
(3.0\,$\pm$\,0.9 mas\,yr$^{-1}$). 
I use Monte Carlo simulations to estimate the observed scatter 
expected in $\mu_{\tau}$ accounting for both the Tycho-2
proper motion errors and the intrinsic velocity dispersion 
of the association.
The intrinsic velocity dispersion $\sigma^{\star}_{int}$ (in km\,s$^{-1}$) 
translates into a proper motion dispersion 
$\sigma(\mu_{\tau}^{int})$ (in mas\,yr$^{-1}$)
as a function of mean subgroup parallax $\pi$ 
(adapted from de Bruijne (1999) eqn. 20):

\begin{equation}
\sigma(\mu_{\tau}^{int}) = \pi\,\sigma^{\star}_{int}/A
\end{equation}

\noindent where A = 4.74 km\,s$^{-1}$\,yr$^{-1}$.
For the simulations, I model velocity dispersions ranging from 
$\sigma^{\star}_{int}$ = 
0-3 km\,s$^{-1}$ in 0.5 km\,s$^{-1}$ steps.
I adopt the mean distances to LCC and UCL 
from de Zeeuw et al. (1999).
I generate 10$^{4}$ Gaussian deviates for each star with zero mean and 
a standard deviation equal to the square root of the observed
value of $\sigma$($\mu_{\tau}$) and the model 
$\sigma$($\mu_{\tau}^{int}$) values added in quadrature. Statistical
testing showed that clipping the Monte Carlo data at the box
boundaries in Fig. 1 ($\left|\mu_{\tau}\right|$ $<$ 8 mas\,yr$^{-1}$) had 
negligible effect on the probability plot determinations of 
 $\sigma$($\mu_{\tau}$), so all Monte Carlo values were retained.\\ 

A comparison between the $\sigma$($\mu_{\tau}$) values for
the Monte Carlo simulations and the observations is shown in Table 1.
It appears that the internal velocity dispersions of the subgroups
are indeed detectable. 
The observed $\sigma$($\mu_{\tau}$) values for LCC and UCL are
consistent with internal velocity dispersions of 
$\sigma^{\star}_{int}$ = 0.6$^{+0.5}_{-0.6}$ km\,s$^{-1}$ 
and 1.0$^{+0.6}_{-1.0}$ km\,s$^{-1}$, respectively.
The 95\% confidence level upper limits 
to the velocity dispersions are $<$1.6\,km\,s$^{-1}$ 
(LCC) and $<$2.2\,km\,s$^{-1}$ (UCL).
We can rule out velocity dispersions of 3\,km\,s$^{-1}$
(de Zeeuw et al. 1999), as this would have
produced a dispersion of $\sigma$($\mu_{\tau}$) $\simeq$ 5-6 mas\,yr$^{-1}$
in both subgroups.

\begin{table}[ht]
\caption[Estimates of  $\sigma$($\mu_{\tau}$) from Observations and Monte Carlo Simulations]
{Estimates of  $\sigma$($\mu_{\tau}$) from Monte Carlo Simulations and Observations}
\begin{tabular*}{\textwidth}{@{\extracolsep{\fill}}lcc}
\it                               & LCC $\sigma$($\mu_{\tau}$) & UCL $\sigma$($\mu_{\tau}$)\cr
\it Sample                        & (mas\,yr$^{-1}$)           & (mas\,yr$^{-1}$)          \cr
\sphline
Observed RASS-TYC2 sample  & 2.89\,$\pm$\,0.46 & 3.48\,$\pm$\,0.49\cr
\sphline
Model $\sigma^{\star}_{int}$ = 0.0 km\,s$^{-1}$ & 2.62 & 3.18 \cr
Model $\sigma^{\star}_{int}$ = 0.5 km\,s$^{-1}$ & 2.78 & 3.27 \cr
Model $\sigma^{\star}_{int}$ = 1.0 km\,s$^{-1}$ & 3.24 & 3.48 \cr
Model $\sigma^{\star}_{int}$ = 1.5 km\,s$^{-1}$ & 3.84 & 3.89 \cr
Model $\sigma^{\star}_{int}$ = 2.0 km\,s$^{-1}$ & 4.53 & 4.35 \cr
Model $\sigma^{\star}_{int}$ = 2.5 km\,s$^{-1}$ & 5.30 & 4.86 \cr
Model $\sigma^{\star}_{int}$ = 3.0 km\,s$^{-1}$ & 6.08 & 5.41 \cr
\sphline
\end{tabular*}
\end{table}
The velocity dispersions determined from the Monte Carlo simulations
are strictly upper limits only. 
The MML02 survey observed {\it most} of the stars in the selection
boxes in Fig. 1, and some of those are known not to be
pre-MS members. Interlopers
will evenly populate the Fig. 1 selection boxes, leading to 
slightly inflated dispersions in $\mu_{\tau}$, although the
use of probability plots largely mitigates against this effect. 
Taking into account the lack of spectroscopic confirmation of pre-MS
status for all of the RASS-TYC2 candidate members, 
I conservatively conclude the following:
{\it The intrinsic velocity dispersion 
$\sigma^{\star}_{int}$ of post-T Tauri stars in the LCC and
UCL OB subgroups is $\leq$2\,km\,s$^{-1}$, similar to
that measured for the early-type members (de Bruijne 1999a).}
A more detailed study, including radial velocity data, 
and answering whether the subgroup expansion is detectable, 
is underway (Mamajek, in prep.).

\section{Is There an OB Subgroup in Chamaeleon?}

Sartori et al. (2003; hereafter SLD03) recently presented a membership 
list of 21 B stars in the
Chamaeleon region that they claim constitute a new OB subgroup of Sco-Cen
(\S2.3 and Table 5 of their paper). The putative Cha OB members 
were selected solely by distance (120-220 pc) and projected
proximity to the Chamaeleon molecular clouds.
I present two observations which demonstrate
that either the Cha subgroup membership list of SLD03 is
severely contaminated by field stars, or that the
group doesn't exist.\\

SLD03 measured a velocity dispersion for their Cha B-star sample of
($\sigma_U$, $\sigma_V$, $\sigma_W$) = (8, 11, 6) km\,s$^{-1}$. 
Observations of nearby OB associations show that their velocity dispersions
are small -- typically $\leq$1-3\,km\,s$^{-1}$ (\S 2; 
de Zeeuw et al. 1999; de Bruijne 1999). Torra et al. (2000) find that young 
($<$100 Myr-old), nearby ($d$ = 100-600 pc) field
OB stars {\it distributed all over the sky} have a velocity dispersion of
($\sigma_U$, $\sigma_V$, $\sigma_W$) $\simeq$ (8, 9, 5) km\,s$^{-1}$. This is
similar to the velocity dispersion for the Cha B stars, 
and suggests that {\it if} the sample contains a bona fide OB 
association, it is probably severely contaminated by field B stars.\\

There is also a discrepancy in the numbers of Cha OB members
versus non-members in the Chamaeleon region. 
If one searches the {\it Hipparcos} catalog for B stars in the 
180 deg$^2$ region surveyed for Chamaeleon {\it ROSAT} T Tauri stars
by Alcal\'a et al. (1995), constrained to distances 
between 120-220 pc, one finds that all 12 B stars 
within these constraints are 
considered Cha OB members by SLD03. 
Where are the non-member, field B stars? How many would one expect?
The projected density of 
{\it Hipparcos} B stars with distances of 120-220 pc at 
Galactic latitude --18$^{\circ}$ is $\sim$0.057 deg$^{-2}$ 
(calculated in a 10$^{\circ}$-wide band centered on $b$ = --18$^{\circ}$, 
covering all Galactic longitudes). Over the 180 deg$^{2}$ Cha region
defined by Alcal\'a, we expect to find 10\,$\pm\,\sqrt{10}$ 
B-type field stars. One finds 12,
consistent with the density of field B-stars, 
and within the 1-$\sigma$ Poisson error bar. 
It is difficult to accept that all 12 B-type {\it Hipparcos} stars 
in this region are members of a new OB 
association -- with zero non-member field B stars -- especially
when 10 non-member B stars are predicted to exist. These numbers 
also suggest
that there is not a statistically significant 
over-density of B stars in the Chamaeleon region.
Along with the high velocity dispersion
of the putative Cha OB membership list, the evidence presented here
suggests that there is no OB subgroup in Chamaeleon.

\section{Future Prospects}

Projects are underway to further understand the nature of the 
LCC and UCL post-T Tauri
stars and their circumstellar environs. E. Mamajek, M. Meyer, 
P. Hinz, \& W. Hoffmann have recently
completed a 3-10\,$\mu$m survey for cool accretion disks among 
$\sim$40 of the Sco-Cen 
post-T Tauri stars using the MIRAC/BLINC mid-IR imaging system on the Magellan 
I telescope. It is unclear whether pre-MS stars can retain cooler disks 
(observed at longer $\lambda$) for 
times longer than the observed lifetime for inner
disks traced by JHKL data ($\sim$10 Myr). Further, its unclear
whether these disks can regulate stellar angular momentum evolution. 
Approximately 30 of the LCC and UCL post-T Tauri stars are among the $\sim$350 
young, solar-type stars in the {\it Formation and Evolution of Planetary 
Systems} (FEPS) SIRTF Legacy Project (Meyer et al. 2002).  
A high resolution echelle survey of the post-T Tauri star sample was conducted
in June 2002 on the CTIO 4-m telescope for measuring 
accurate radial and rotational 
velocities (Mamajek et al., in prep.). 
The primary goals are (1) to determine the distribution of stellar 
rotational velocities and study stellar angular momentum evolution
in the post-T Tauri phase, (2) use radial velocities to ensure membership, 
as well as determine the kinematic expansion age of the subgroups, and (3) 
determine whether a spread in Li abundances is present.

\begin{acknowledgments}
EEM is currently supported by a NASA Graduate Student Researchers Program 
fellowship (NGT5-50400) and recently by NASA contract 1224768 
administered by JPL. I thank the meeting organizers for 
allowing me to give an oral presentation, 
and for making the Ouro Preto meeting such an enjoyable success. 
I also thank Mike Meyer and Lissa Miller for
critiquing drafts of this manuscript. 
\end{acknowledgments}

\end{document}